\documentclass[referee]{raa}
\usepackage{graphicx,times}
\usepackage[varg]{txfonts}
\usepackage{natbib}
\bibpunct{(}{)}{;}{a}{}{,}
\begin{document}

\title{AN INVESTIGATION OF A MAGNETIC CATACLYSMIC VARIABLE WITH A PERIOD OF 14.1 ks}

\volnopage{Vol.0 (200x) No.0, 000--000}      
\setcounter{page}{1}          

\author{Song Wang\inst{1,2} \and Yu Bai\inst{1,2} \and Chuan-peng Zhang\inst{2} \and Ji-feng Liu\inst{1,2,3}}

\institute{Key Laboratory of Optical Astronomy, National Astronomical Observatories,
Chinese Academy of Sciences, Beijing 100012, China;
songw@bao.ac.cn, jfliu@bao.ac.cn
\and
National Astronomical Observatories,
Chinese Academy of Sciences, Beijing 100012, China
\and
College of Astronomy and Space Sciences,
University of Chinese Academy of Sciences, Beijing 100049, China}

\abstract
{Magnetic cataclysmic variables (CVs) contain a white dwarf with magnetic field strong enough to control the accretion flow
from a late type secondary.
In this paper, we discover a magnetic CV (CXOGSG J215544.4+380116) from the $Chandra$ archive data.
The X-ray light curves show a significant period of 14.1 ks, and the X-ray spectra
can be described by a multi-temperature hot thermal plasma, suggesting the source as a magnetic CV.
The broad dip in the X-ray light curve is due to the eclipse of the primary magnetic pole,
and the additional dip in the bright phase of the soft and medium bands
may be caused by the accretion stream crossing our line of sight to the primary pole.
Follow-up optical spectra show features of an M2--M4 dwarf dominating the red band and
a WD which is responsible for the weak upturn in the blue band.
The mass ($\sim$ 0.4 $M_{\odot}$) and radius ($\sim$ 0.4 $R_{\odot}$) for the M dwarf are obtained using
CV evolution models and empirical relations between the orbital period and the mass/radius.
The estimated low X-ray luminosity and accretion rate may suggest the source as a low-accretion-rate polar.
In addition, Very Large Array observations reveal a possible radio counterpart to the X-ray source, but with a low significance.
Further radio observations with high quality are needed
to confirm the radio counterpart and explore the properties of this binary system.
\keywords{novae, cataclysmic variables -- Stars: late-type -- Stars: magnetic field}
}

\authorrunning{Wang et al.}
\titlerunning{MAGNETIC CATACLYSMIC VARIABLE}
\maketitle

\section{Introduction}
\label{intro.sec}

Cataclysmic variables (CVs) are close binaries including
a white dwarf (WD) primary accreting matter from a late-type, Roche lobe-filling
donor star \citep{Warner1995a}.
One special subclass, representing around 20\% of known CVs, is magnetic CVs.
Their magnetic field is strong enough to
control at least the inner part of the accretion flow.
The magnetic CVs are divided into two classes, namely polars,
where field strength is high enough to lock the WD rotation to
the binary orbit and prevent the formation of an accretion disk,
and intermediate polars (IPs),
where WDs have weaker magnetic field so that an outer disk can form \citep{Martino2008}.
Therefore, magnetic CVs are important objects for
understanding accretion processes in the presence of a strong magnetic field \citep{Schwarz2009}.

$Chandra$ X-ray Observatory is one of the most successful X-ray missions.
With the unprecedented subarcsecond spatial resolution
(e.g., 10 times superior to that of $ROSAT$ HRI) and low sensitivity limit,
$Chandra$ provides a unique view of the X-ray sky 10--100 times deeper than previously \citep{Weisskopf2000}.
The $Chandra$ observations have also been proved powerful in hunting for
coherent or quasi-coherent signals \citep{Esposito2013a, Esposito2013b}.
In this paper, we present a detailed study of one X-ray
point source CXOGSG J215544.4+380116 (hereafter CXOGSG J2155),
the X-ray periodicity of which was serendipitously discovered
when we reduced and analyzed the $Chandra$ archive data \citep{Wang2016}.
We describe time analysis and X-ray spectra study in Section \ref{xdata.sec}.
To examine the nature of CXOGSG J2155, we
collect optical observations, and obtain spectroscopic observation
with the 2.4 m telescope at the Lijiang Observatory.
We present optical data analysis and results in Section \ref{odata.sec},
and describe radio observations in Section \ref{rdata.sec}.
A discussion about the nature of the binary follows in Section \ref{discuss.sec}.

\section{X-ray observations and data analysis}
\label{xdata.sec}

CXOGSG J2155 was observed serendipitously by {\it Chandra} three times (Obsid 3967, 13218, and 12879)
in observations targeting the seyfert galaxy 3C 438.
The data were acquired with the Advanced CCD Imaging Spectrometer (ACIS)
in Very Faint imaging (Timed Exposure) mode, with time resolution around 3.2 s.
See Table \ref{chandra.tab} for a summary of the observations.
All these observations were reprocessed with the $Chandra$ Interactive Analysis of Observations software
(CIAO, version 4.6).

\subsection{Counts and Colors}

The {\tt wavdetect} tool is used to detect the source \citep{Freeman2002}
and derive the $3\sigma$ elliptical source region.
The background-subtracted photon counts with its error from {\tt wavdetect} are
161.0 $\pm$ 13.0, 154.0 $\pm$ 12.7, and 246.0 $\pm$ 16.1 for the three observations, respectively.

The source event lists are extracted from the $3\sigma$ elliptical source region.
The vignetting corrected count rate in 0.3--8 keV is 3.39, 3.41, and 3.25 ks$^{-1}$,
so no significant long-term variability is found.
Three energy bins are employed to define hardness ratio following \citet{Di Stefano2003a}:
soft ($S$: 0.1--1.1 keV), medium ($M$: 1.1--2 keV), and hard ($H$: 2--7 keV).
With the hierarchical classification scheme \citep{Di Stefano2003b}, CXOGSG J2155 is identified as a hard source
in all the three observations.
Table \ref{chandra.tab} lists the key parameters for each observation, including
Obsid, exposure time, observation date, off-axis angle, vignetting factor, detection significance,
background-subtracted photon counts with error, expected background counts in the source region, X-ray colors, and
source classification.

\begin{table}
\begin{center}
\caption[]{Key Parameters for Each $Chandra$ Observation of CXOGSG J2155.}
\label{chandra.tab}
\begin{tabular}{lcccccccccc}
\hline\noalign{\smallskip}
ObsID  &    Expos.  &    DATE  &      OAA          &  VigF  & $\sigma$  &  Counts  &  bkgd  &  ${M-S \over H+M+S}$  &  ${H-M \over H+M+S}$  \\
       &    (sec)   &            & ($^{\prime\prime}$)  &        &           &          &        &           &                \\
  (1)  &     (2)    &       (3)  &       (4)         &   (5)  &      (6)  &     (7)  &    (8) &      (9)  &      (10)         \\
\hline\noalign{\tiny}
acis3967 & 47467 & 2002/12/27 & 147 & 0.832 & 43.6 & 161.0 (13.0) & 1.3 & 0.32 (0.15) & -0.11 (0.14) \\
acis13218 & 47442 & 2011/01/28 & 182 & 0.967 & 38.6 & 154.0 (12.7) & 1.3 & 0.27 (0.16) & -0.02 (0.15) \\
acis12879 & 72036 & 2011/01/30 & 182 & 0.970 & 51.0 & 246.0 (16.1) & 1.9 & 0.34 (0.11) & -0.27 (0.11) \\
\hline\noalign{\smallskip}
\end{tabular}
\end{center}
\tablecomments{1\textwidth}{The columns are:
(1) observation ID;
(2) source exposure time after deadtime correction;
(3) observation date;
(4) off-axis angle in arcsecond;
(5) vignetting factor calculated from the exposure map being the ratio between the local and the maximum map value;
(6) detection significance from {\tt wavdetect};
(7) background-subtracted photon counts with its uncertainty in parenthesis;
(8) background counts within the source region;
(9) X-ray color $C_{MS} = (M-S)/(H+M+S)$ with uncertainty in parenthesis. S/M/H represents background-subtracted counts in soft (0.3-1 keV), medium (1-2 keV), and hard (2-8 keV) bands;
(10) X-ray color $C_{HM} = (H-M)/(H+M+S)$ with uncertainty in parenthesis.}
\end{table}

\subsection{Timing Analysis}
\label{xtime.sec}

\begin{figure*}[!htb]
\center
\includegraphics[width=0.8\textwidth]{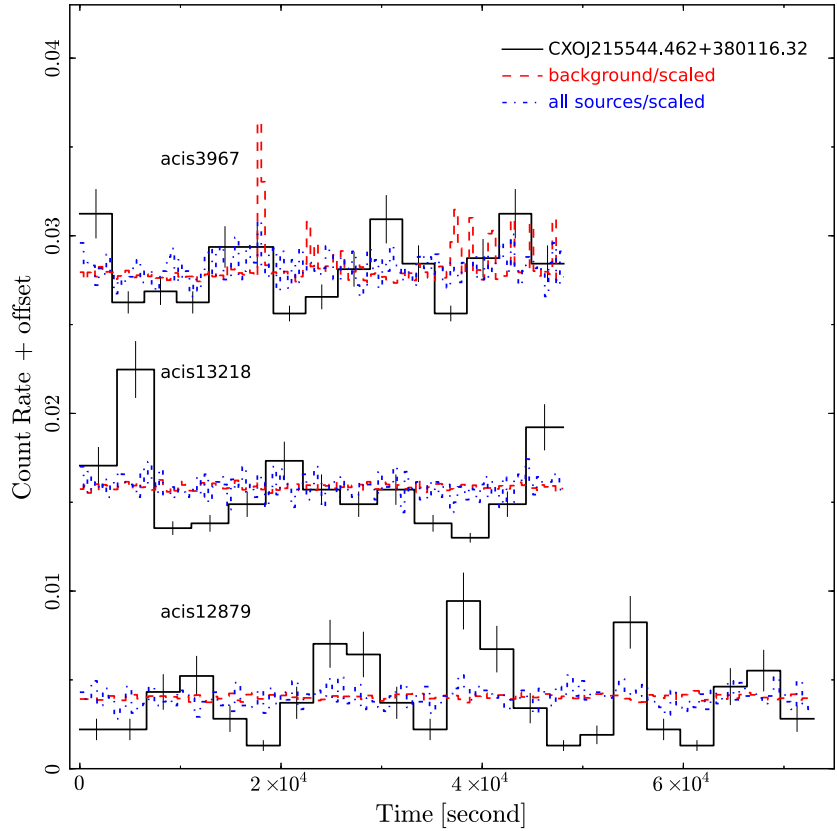}
\caption{Binned light curve for CXOGSG J2155 in $Chandra$ ACIS observations.
The blue and red binned light curves are for all detected sources and for the total
background of S3 chip overplotted for comparison,
which clearly shows that these eclipses are not caused by detector background or
flares but a behavior unique to this source itself.}
\label{allmlc.fig}
\end{figure*}

\begin{figure*}[!htb]
\center
\includegraphics[width=0.48\textwidth]{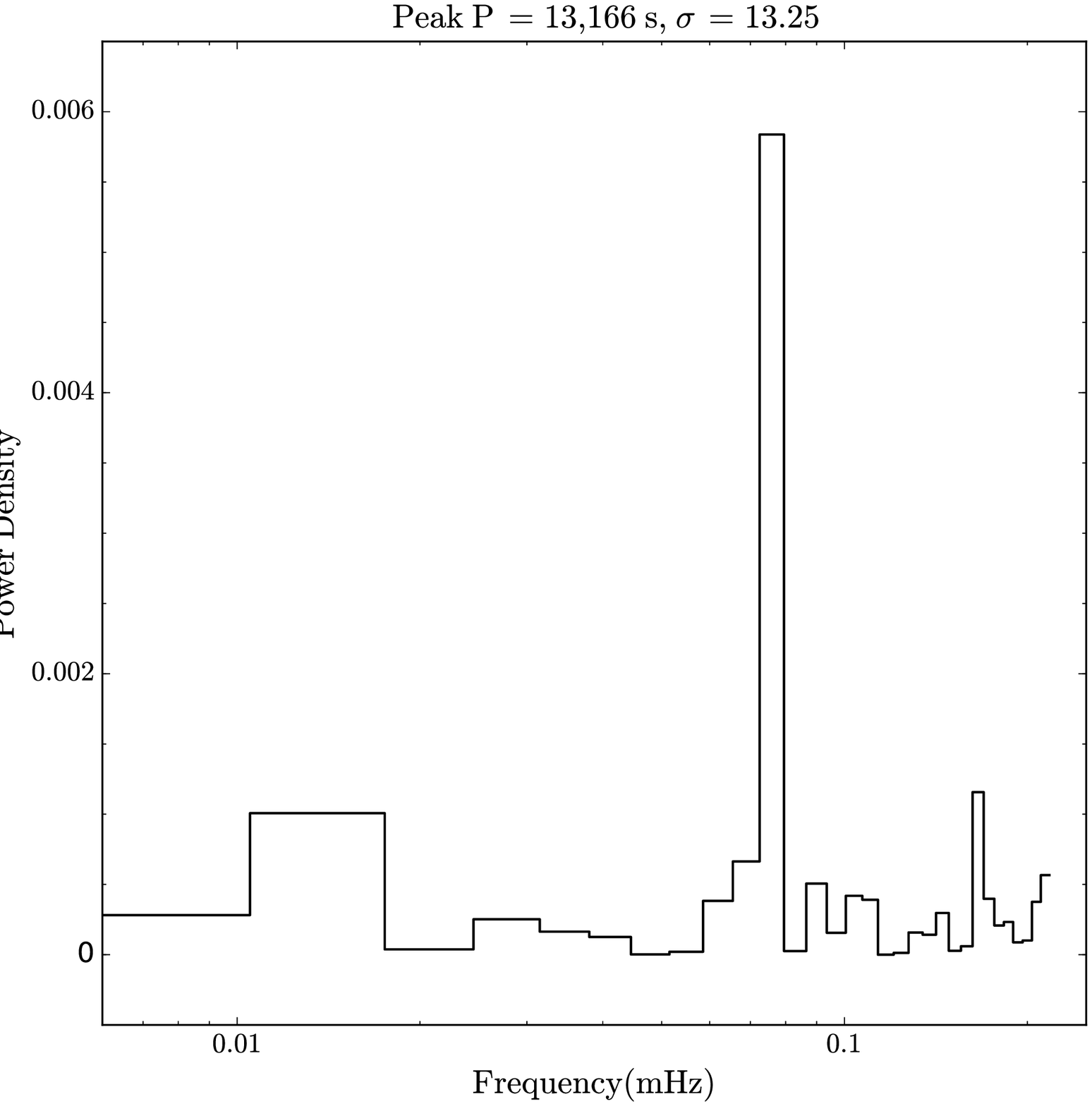}
\hspace{1 mm}
\includegraphics[width=0.48\textwidth]{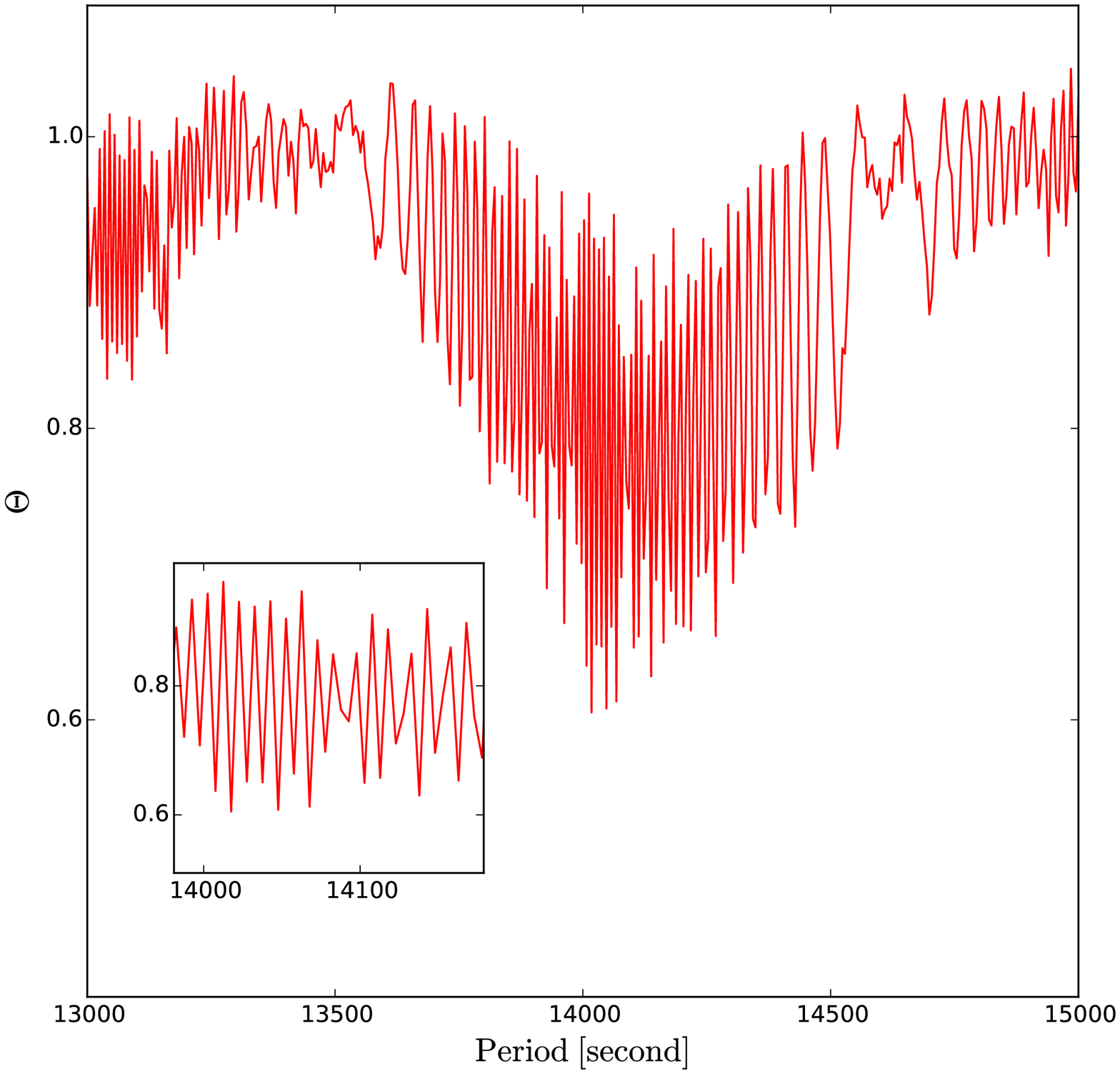}
\caption{Left panel: Power spectra of CXOGSG J2155 obtained from $Chandra$ ACIS observation (Obsid 12879).
A peak at frequency $\nu$ = 1/P $\simeq$ 0.076 mHz (P $\simeq$ 13.17 $\pm$ 1.21 ks) can be clearly seen.
Right panel: The phase dispersion calculated with the three $Chandra$ light curves,
for the periods from 13 ks to 15 ks, with
a step of 5 s. The insert shows several phase dispersion minima.}
\label{frequency.fig}
\end{figure*}

The {\tt axbary} tool is applied to all observations for barycenter correction.
The light curves of CXOGSG J2155 displays notable pulsed emission (Figure \ref{allmlc.fig}),
and we applied two techniques to search for possible periods. The first is the
fast fourier transform (FFT) method. The fourier power spectrum is computed with the data set of Obsid 12879,
which has the largest count and longest exposure and can be used to perform an effective examination.
A possible signal is found in the power spectrum (Figure \ref{frequency.fig}, left), and the high peak shows
an approximate period of $\sim$ 13.17 $\pm$ 1.21 ks ($\nu$ $\simeq$ 0.076 mHz).
The significance is computed as $\sigma = 13.25$, which is
the ratio of the power density of the peak to the averaged power density.
Another method applied to the data is the phase dispersion minimization (PDM) analysis\citep{Stellingwerf1978}.
A search of periods in the range of 13--15 ks with steps of 5 s
reports a group of phase dispersion minima at
14,007.52, 14,017.54, 14,027.57, 14,037.60, 14,047.62, 14,067.67, 14,102.76, 14,137.84,
and 14,162.91 s (Figure \ref{frequency.fig}, right).
Searches of periods are then performed with finer steps of 0.01 s around these minima ($\pm$ 5 s).

\begin{figure*}[!htb]
\center
\includegraphics[width=0.48\textwidth]{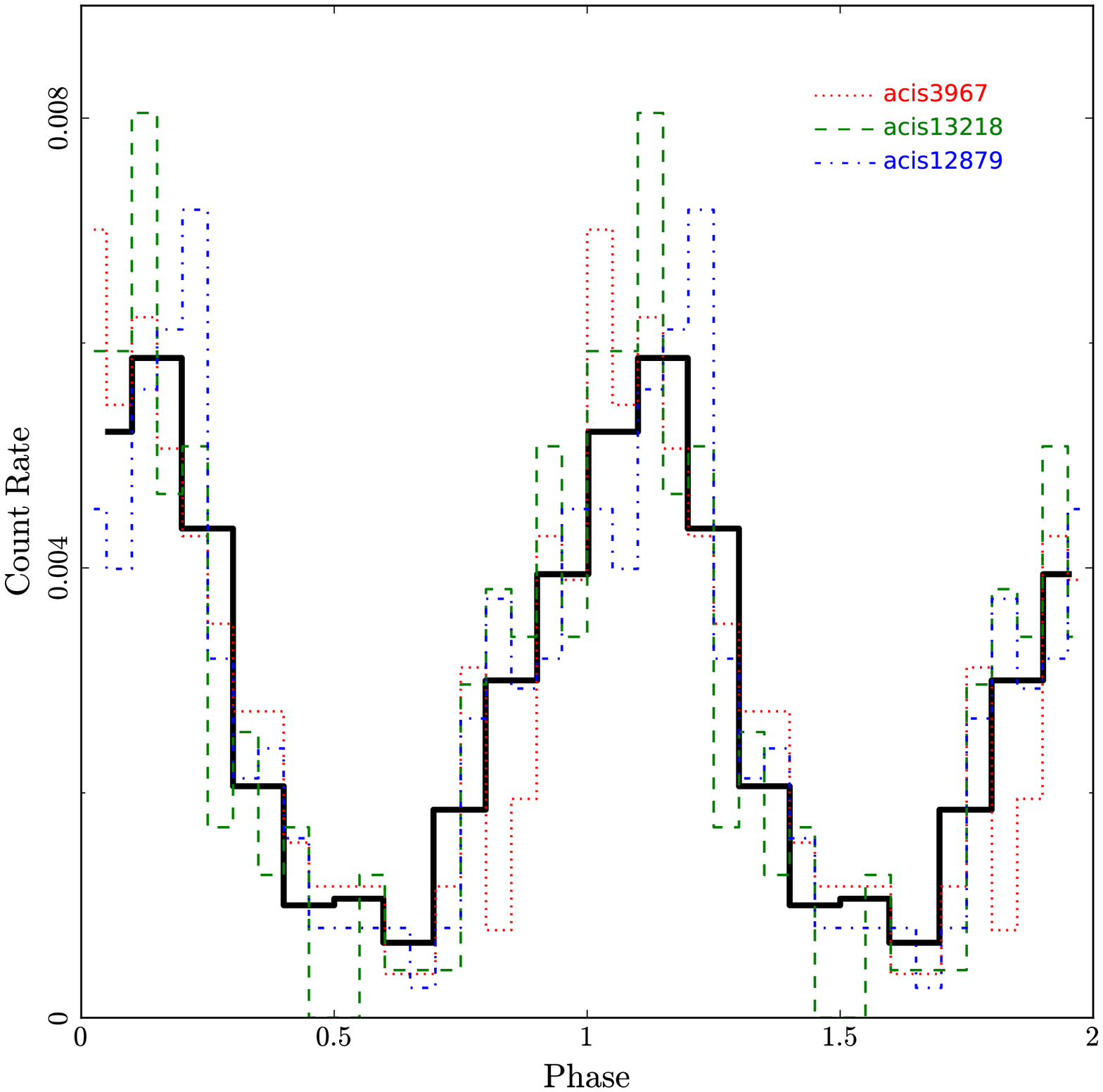}
\hspace{1 mm}
\includegraphics[width=0.48\textwidth]{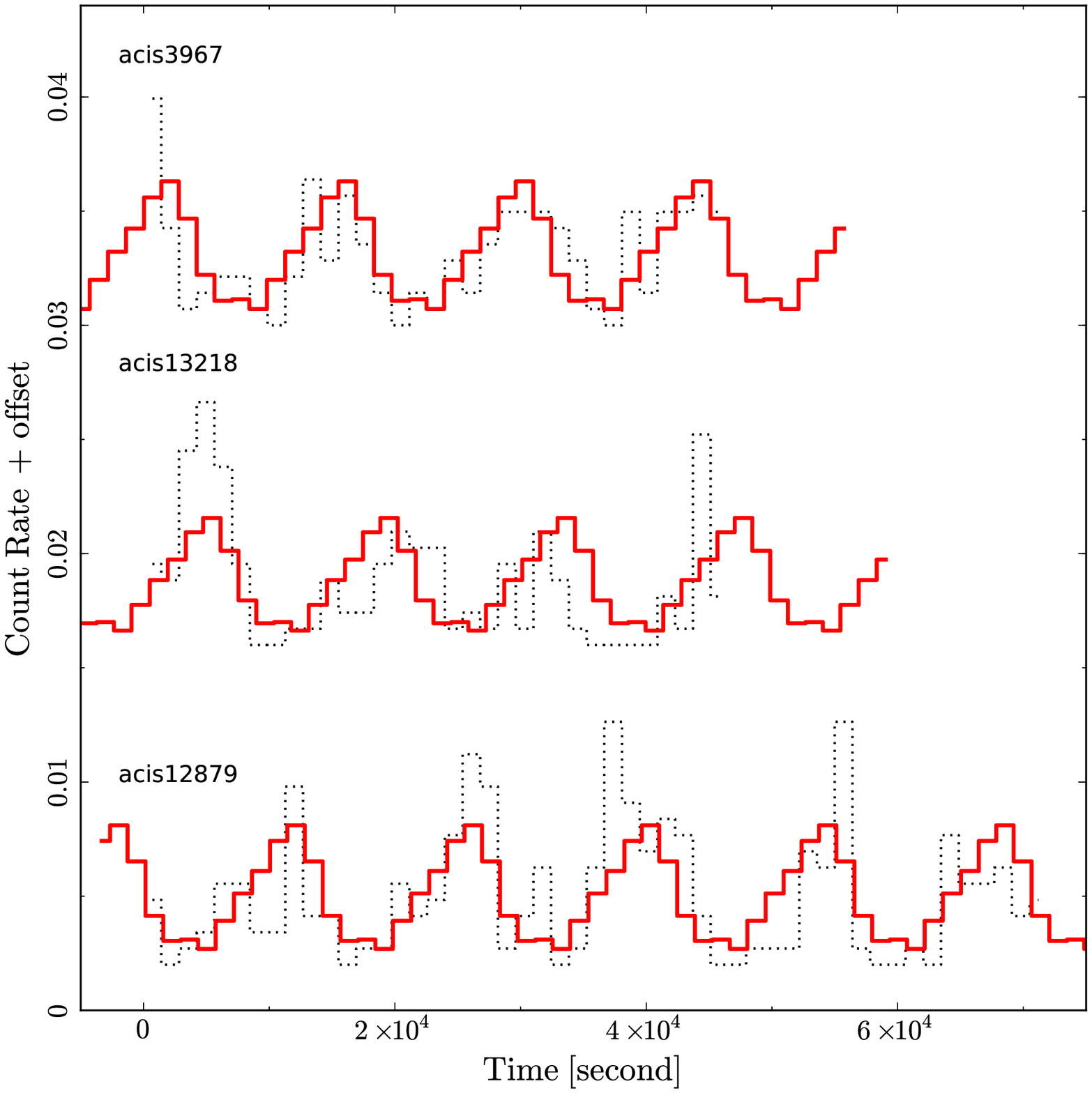}
\caption{Left panel: Light curves folded with the period of 14101.10 s for the three observations.
The black solid histogram is the average light curve from the three observations.
Right panel: Individual light curves overplotted to the
simulated average light curve for the period of 14101.10 s.}
\label{14101.1com.fig}
\end{figure*}

With these period candidates determined from the FFT and PDM methods,
we fold the light curves with each possible period to compute an averaged light curve,
and compare them to individual light curves to check the phase preservation \citep{Liu2006}.
The Pearson product-moment correlation coefficients are computed for these possible periods.
The average light curve folded with a period of 14101.10 s (Figure \ref{14101.1com.fig}, left) show a remarkable correlation with
the individual light curves, with the largest correlation coefficient as 0.96.
The light curves at the time windows for the observations are then simulated using the period of 14101.10 s,
which display great consistency in the phase and the flux level with individual light curves
(Figure \ref{14101.1com.fig}, right).
The time line for the three observations covers a range of eight years,
corresponding to $\sim$ 11,384 cycles.
With the well preserved phase (better than 1/10),
the fractional error in the period is $\sim$ (1/10)/11384 $\sim$ 8.8$\times10^{-6}$.
Therefore, we conclude the period is 14101.10 $\pm$ 0.12 s.
Recently, \citet{Israel2016} reported the discovery of 41 pulsating sources using $Chandra$ observations,
showing CXOGSG J2155 with a period of 14090 $\pm$ 43 s,
in good agreement with our result.

\subsection{X-ray Spectral Analysis}
\label{xspec.sec}

Source spectra are extracted from the $3\sigma$ elliptical source region with {\tt specextract}.
The corresponding background spectrum is extracted from its local background region,
which is an annulus immediately surrounding the source, with internal (external) radius of 2 (4) fold
of the long-axis of the source region.

\begin{figure*}[!htb]
\center
\includegraphics[width=0.8\textwidth]{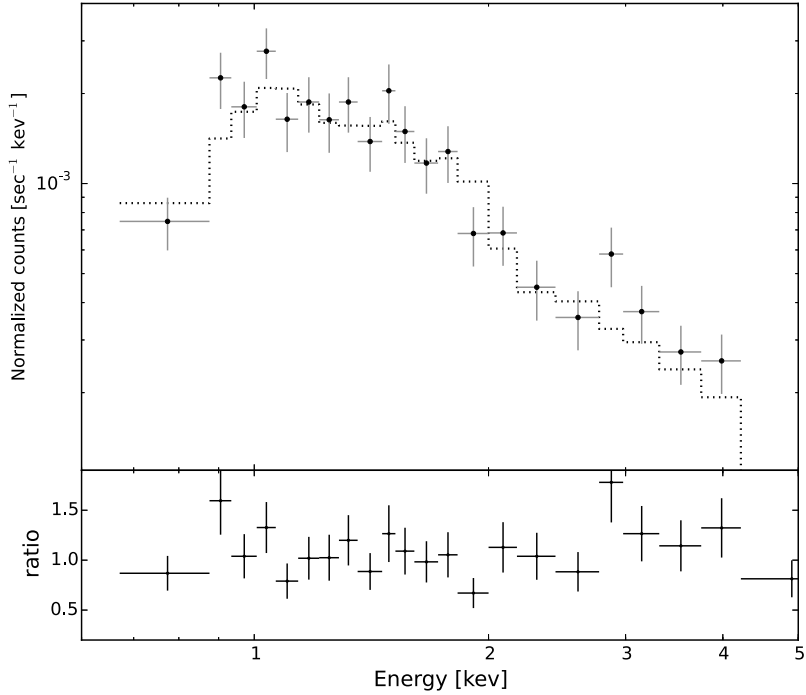}
\caption{The combined spectrum (0.3--8$~$keV) from three observations is shown with black crosses,
and the best-fitting model [PHABS(CEMEKL)] is shown with black dotted lines.
The error bars denote 68.3\% uncertainty.}
\label{spec.fig}
\end{figure*}

The spectral fitting is performed with the {\tt xspec} package \citep{Arnaud1996} for
the merged spectrum of three observations (Figure \ref{spec.fig}).
Three models are used for the fitting: power law, bremsstrahlung, and cemekl,
all modified for the interstellar absorption.
The cemekl model is a multitemperature version of the mekal code,
with the emission measure defined as a power law in temperature \citep{Singh1996}.
The fitting results are summarized in Table \ref{xspec.tab}.
The power-law photon index is rather flat ($\Gamma = 1.85 \pm 0.18$), indicating a hard spectrum.
It is clear that the cemekl code, the multi-temperature plasma emission model,
has better fitting results (smaller $\chi^{2}$) than the power law and bremsstrahlung model.
The fitted equivalent absorbing column is $N_H = (3.5\pm0.8) \times 10^{21}$ cm$^{-2}$,
corresponding to a color excess as $E(B-V) \simeq 0.51\pm0.12$ with $N_H = 6.86\times10^{21} E(B-V)$ \citep{Guver2009}.

\begin{table}
\begin{center}
\caption[]{X-ray Spectral Fitting Results of CXOGSG J2155.}
\label{xspec.tab}
\begin{tabular}{lcccccc}
\hline\noalign{\smallskip}
Model &   $N_H$             & $\Gamma$   & $KT^{a}$  &    Flux & Unabsorbed Flux & $\chi^2_{\nu}$/dof \\
      & (10$^{22}$ cm$^{-2}$) &  &  (keV) & \multicolumn{2}{c}{(10$^{-14}$ erg cm$^{-2}$ s$^{-1}$)}  & \\
  (1)  &     (2)    &       (3)  &       (4)         &   (5)  &      (6)  &     (7)  \\
\hline\noalign{\tiny}
PHABS(POWERLAW)    & 0.31$\pm$0.07 & 1.85$\pm$0.18 &                & 3.93E-14 & 6.02E-14 & 1.367/19  \\
PHABS(BREMSS)      & 0.23$\pm$0.06 &               & 5.76$\pm$1.98  & 3.79E-14 & 5.00E-14 & 1.367/19 \\
PHABS(CEMEKL)      & 0.35$\pm$0.08 &               & 13.39$\pm$5.23 & 4.13E-14 & 5.95E-14 & 1.128/19  \\
\hline\noalign{\smallskip}
\end{tabular}
\end{center}
{$^{\rm a}$}{For the cemekl model, $T$ means the $T_{\rm max}$ parameter.}
\end{table}

\section{Optical data analysis}
\label{odata.sec}

\subsection{Astrometry and Photometry}
\label{oastro.sec}

\begin{figure*}[!htb]
\center
\includegraphics[width=0.8\textwidth]{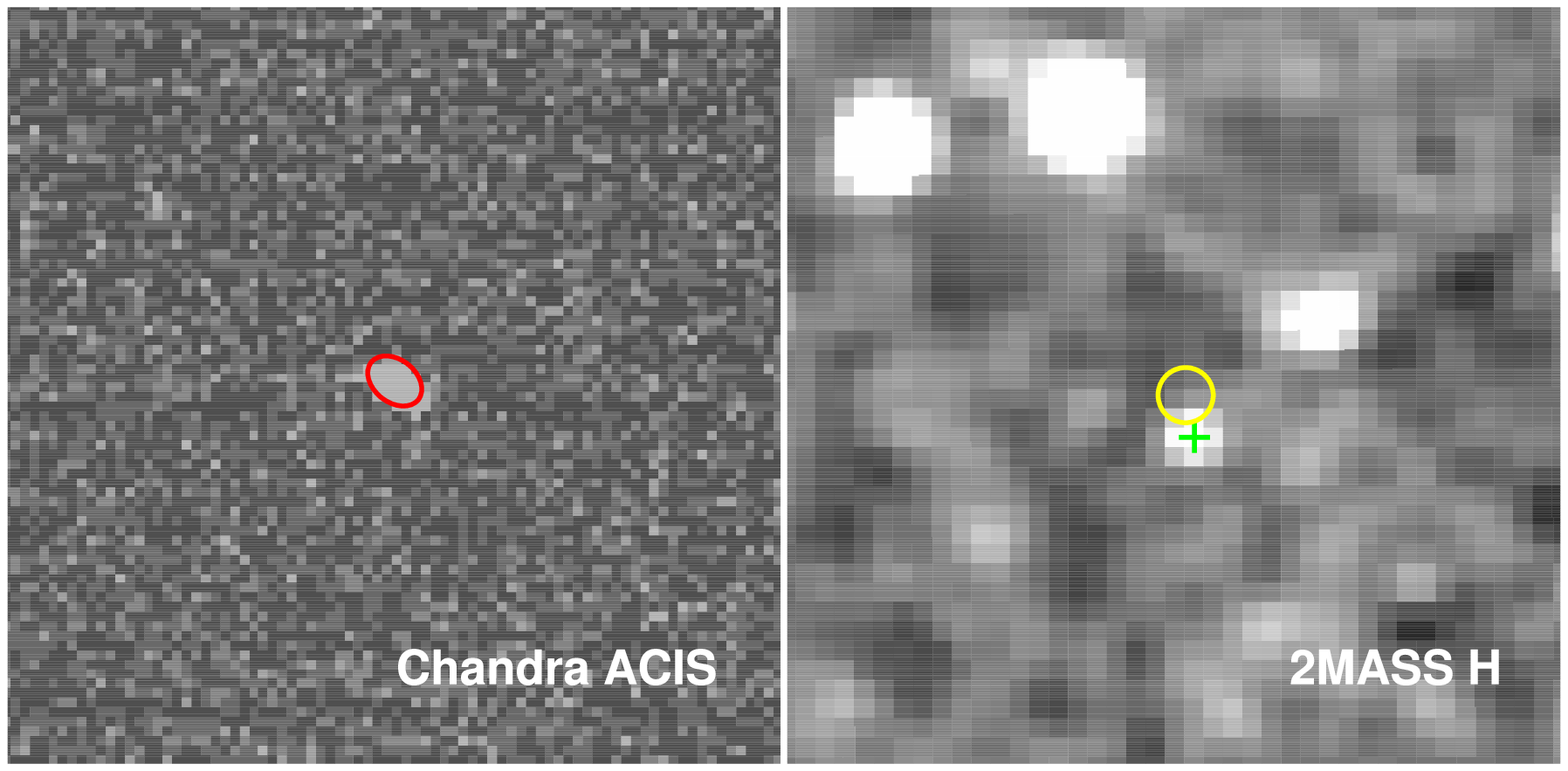}
\caption{Left panel: The $Chandra$ ACIS image around CXOGSG J2155. The ellipse shows
the X-ray position determined from {\tt wavdetect}.
Right Panel: The 2MASS image around CXOGSG J2155 in $H$ band.
The yellow circle ($r = 1.4''$) represents the corrected X-ray position.}
\label{2mass.fig}
\end{figure*}

There is an optical object located close to the X-ray position of CXOGSG J2155.
To examine whether this object is the true counterpart, we register the $Chandra$ images onto the 2MASS frame
and compute the error circle for CXOGSG J2155.
This is achieved by identifying four 2MASS sources on the $Chandra$ image.
With the help of the multiple matched pairs of objects,
we calculate the correction used to translate the {\it Chandra} position of CXOGSG J2155  onto the 2MASS image as
$-0.72''\pm0.74''$ in right ascension and $-0.73''\pm0.83''$ in declination.
An error circle is determined with the radius as a quadratic
sum of the astrometric errors from the $Chandra$-2MASS registration
and the position error ($1''$) of CXOGSG J2155 from Chandra observations.
As shown in Figure \ref{2mass.fig} (right), the optical object (2MASS 21554436+3801135) is located close to the error circle, and
we consider it as the true counterpart.

We select optical photometry from archival data.
The USNO A2.0 present magnitudes for CXOGSG J2155 with $B=18.6$ and $R=17.4$ \citep{Monet1998},
which can be transformed to Johnson $UBVRI$ magnitudes using the relationships between these two
systems derived by \citet{Fabregat2005}.
Finally, the Johnson $UBVRI$ magnitudes for CXOGSG J2155 are calculated as $U=19.31\pm0.46$,
$B=18.80\pm0.41$, $V=17.93\pm0.26$, $R=17.37\pm0.32$, and $I=16.85\pm0.26$.
Because no magnitude error is provided in the USNO A2.0 catalog,
the photometry uncertainty only comes from the transformation between the two magnitude scales.
The 2MASS also report magnitudes for CXOGSG J2155, as
$J=16.661\pm0.134$, $H=15.652\pm0.124$, and $K=15.828$ (upper limit).

\subsection{Optical Spectral Analysis}
\label{ospec.sec}

In order to confirm the stellar type of the companion,
the 2.4 m telescope at Lijiang Observatory was used to obtain its optical spectrum.
The optical spectra were taken on Dec 01, 2015 with the Yunnan Faint Object Spectrograph and Camera,
and a G8 grism was used in the observation, with a long slit approximate 1.8$''$ in width.
This source was observed twice, each for 2400 s exposures.
The IRAF package is applied to perform the standard CCD procedures including
overscan and bias subtraction, flat correction, cosmic-ray removal, spectrum extraction, and wavelength
calibration using He/Ne lamps.

\begin{figure*}[!htb]
\center
\includegraphics[width=0.8\textwidth]{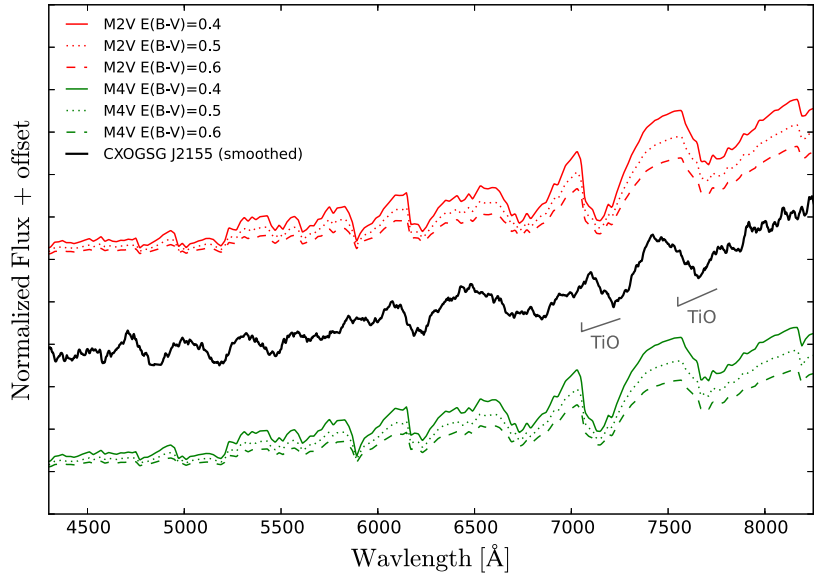}
\caption{Comparison of the merged and smoothed CXOGSG J2155 spectrum with model spectra from M2V and M4V.}
\label{optspec.fig}
\end{figure*}

Due to the low signal to noise ratio of the spectrum, no emission/absorption line could be identified.
However, \citet{Israel2016} simply mentioned that
their spectroscopic follow-up observation of the optical counterpart of CXOGSG J2155 shows H and He emission lines.
Figure \ref{optspec.fig} displays the merged and smoothed CXOGSG J2155 spectrum.
It shows clear features of an M dwarf dominating the red band and a WD which is responsible for the weak upturn in the blue band,
similar to the pre-polar WX LMi \citep{vogel2007} and the MCV SDSS J2048+0050 in low accretion-rate states \citep{Schmidt2005}.
From the shape of the TiO bands at 7053--7861 {\AA} and the shape of the continuum reward of 8200 {\AA} \citep{Bhalerao2010},
the red spectral range of CXOGSG J2155 is inferred to be M2--M4.
The model M2/M4 dwarf spectra \citep{Castelli2004}, with different extinction, are plotted for a comparison.
Finally, the Pearson product-moment correlation coefficients are computed for the M2 and M4 models, with the extinction $E(B-V)$
ranging from 0.4 to 0.6. All these models are acceptable, with the correlation coefficient being 0.88 to 0.9.

\section{Radio data analysis}
\label{rdata.sec}

\begin{figure*}[!htb]
\center
\includegraphics[width=0.8\textwidth]{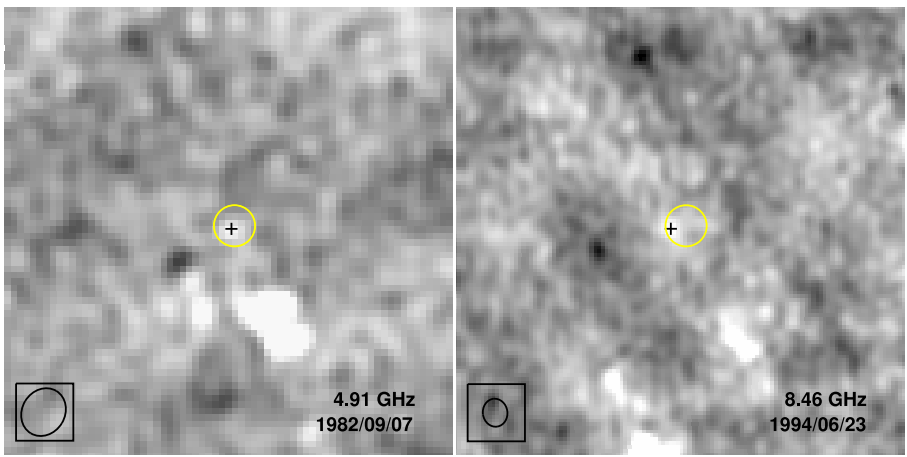}
\caption{The VLA image of 4.91 GHz and 8.46 GHz around CXOGSG J2155.
The yellow circles ($r = 1.04''$)represent the corrected X-ray position.
The synthesized beams are shown by the black ellipses in the bottom left of each panel.}
\label{vla.fig}
\end{figure*}

There is no radio counterpart in NVSS, FIRST, and SUMSS catalogues.
However, when we register the $Chandra$ images onto the VLA frame,
there seems to be one object located close to CXOGSG J2155.
The astrometry is done using the observation with highest spacial resolution (e.g., smallest beam size ),
which was performed
on Jun. 23, 1994, at 8.46 GHz (X-band).
With two $Chandra$ sources identified on the VLA image,
we determine the correction to translate the {\it Chandra} position of CXOGSG J2155  onto the VLA image as
$-0.02''\pm0.13''$ in right ascension and $-0.31''\pm0.25''$ in declination.
An error circle is determined with the radius as a quadratic
sum of the astrometric errors from the $Chandra$-VLA registration
and the position error ($1''$) of CXOGSG J2155 from Chandra observations.
Finally, we find there is an object located within the error circle in two VLA images (Figure \ref{vla.fig}),
and it may be the radio counterpart of CXOGSG J2155.

Three VLA observations at different dates are used to determine the flux density (Table \ref{radio.tab}).
The unit of the VLA data is Jy/beam, which needs to be converted to Jy/pixel before extracting the flux.
This is achieved by multiplying the conversion factor ${\rm 4ln2 \over \pi(beam/pixel)^2}$ \citep{van Hoof2010},
where the pixel and synthesized sizes are about 0.189$''$ and 0.66$''$, respectively.
Aperture photometry for the radio companion is performed with an aperture radius of $1.5''$,
and an annulus immediately surrounding each source,
with inner radius as $2''$ and outer radius as $3''$,
is adopted to obtain the background intensity.
No object is detected in the image of 8.45 GHz, and the upper limit is presented
using the flux in the $1.5''$ circle.
The net radio fluxes after primary beam correction are listed in Table \ref{radio.tab}.

\begin{table}
\begin{center}
\caption[]{Radio Observations of CXOGSG J2155.}
\label{radio.tab}
\begin{tabular}{ccccc}
\hline\noalign{\smallskip}
Frequency  &   DATE &   Flux density &    Pixel size   & Beam size  \\
  (GHz)    &          &   (mJy)        &    ($''$)       & ($''$)     \\
  (1)  &     (2)    &       (3)  &       (4)         &   (5) \\
\hline\noalign{\tiny}
4.91       &  1982/09/07   &   1.57         &    0.33         & 1.16  \\
8.45       &  1993/03/22   &   $<$ 0.07     &    0.19         & 0.65   \\
8.46       &  1994/06/23   &   2.63         &    0.19         & 0.66       \\
\hline\noalign{\smallskip}
\end{tabular}
\end{center}
\end{table}

\section{Discussion}
\label{discuss.sec}

1. $What\ is\ ascribed\ to\ the\ dips\ in\ the\ X$-$ray\ light\ curves?$

Generally, the light curve dips may caused by the accretion stream absorption, a self-eclipse of the accretion region,
or an eclipse by the secondary star.
To examine the mechanism of the dips, we should first to investigate the nature of the compact object.
The X-ray spectra can be described by a multi-temperature hot thermal plasma
(possibly including a complex absorption pattern), suggesting the source as an MCV \citep{Rana2005} rather than a
low-mass X-ray binary.
One striking feature is the lack of a soft X-ray component for CXOGSG J2155,
which is similar to the polar CE Gru \citep{Ramsay2002} and CG X-1 \citep{Weisskopf2004},
although recently the latter one has been reported as a black hole \citep{Esposito2015}.
Using the two-pole emission model,
\citet{Weisskopf2004} reported that the soft X-ray component--the blackbody radiation (10--100 eV) from the soft pole--may be absorbed
by the HI column, leaving only the bremsstrahlung emission ($\sim$ 10 keV) from the secondary hard pole being detected.

\begin{figure*}[!htb]
\center
\includegraphics[width=0.6\textwidth]{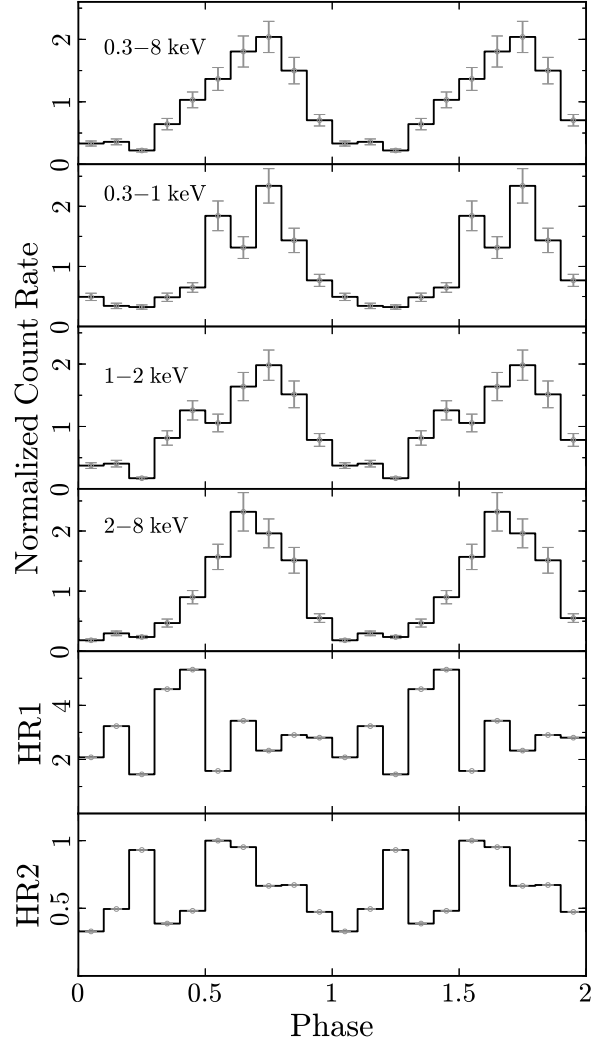}
\caption{Background-subtracted folded pulse profiles in different energy bands.
The hardness ratio HR1 and HR2 are also plotted at the bottom.}
\label{allfold.fig}
\end{figure*}

The folded light curves in different bands are shown in Figure \ref{allfold.fig}.
In addition, two hardness ratio curves are defined and plotted as a function of the phase,
where HR1 is the ratio of the count rate in 1--2 keV to that in 0.3--1 keV energy bands,
and HR2 is the ratio of the count rate in 2--8 keV to that in 1--2 keV energy bands.
The eclipse duration is about half of the period ($\sim$ 1.2 hr), corresponding to
an eclipse full angle of $\sim$ 100$^{\circ}$.
It is distinctly asymmetric, with a gradual rise and a more rapid decline.
The two distinct parts in the X-ray light curve is similar to the polar CE Gru \citep{Ramsay2002},
which can be explained with the two-pole emission model.
The bright phase (0.3--1) X-ray emission is from the primary accreting pole, which
can is visible for seventy percent of the period due to self-eclipse or eclipse by a secondary star;
the faint phase (0--0.3) X-ray emission is from the secondary pole in the other hemisphere,
and the faint X-ray region can be seen for all the phase.
There seems one additional dip in the bright phase of soft/medium band,
which is characteristic of photo-electric absorption
and has been seen in both polars \citep{Watson1989} and IPs \citep{Martino2001, Martino2004}.
The ballistic stream or the stream to the secondary faint pole may cross
our line of sight to the primary magnetic pole, thereby absorbing the soft/medium X-rays and causing the additional dip \citep {Ramsay2002}.
This is not surprising, because these more extended, cooler parts of the accretion region, detected in the soft/medium band,
is more likely to be absorbed,
while the harder emission region may only be absorbed by the dense core of the accretion stream \citep{Traulsen2014}.







2. ${What\ are\ the\ proeprties\ of\ the\ binaries?}$

The period 14.1 ks discovered in the X-ray light curve should be the orbital period,
which locates the binary system above the 2--3 hour so-called ``orbital period gap''.
Many previous studies have reported a relation between the orbital period and the
mass/radius of the secondary star.
\citet{Warner1995b} derived following semi-empirical relations as
$M_2 = 0.065 P_{\rm hr}^{5/4}~ M_{\odot}$
and
$R_2 = 0.094 P_{\rm hr}^{13/12}~ R_{\odot}$.
\citet{Smith1998} obtained new relations using a carefully selected sample of CVs with well-measured
system parameters:
$M_2 = (0.038 \pm 0.003) P_{\rm hr}^{1.58\pm0.09}~ M_{\odot}$
and
$
R_2 = (0.081 \pm 0.019) P_{\rm hr}^{1.18\pm0.04}~ R_{\odot}$.
\citet{Frank2002} also presented the period-mass relation
$M_2 \cong 0.11 P_{\rm hr}~ M_{\odot}$
and the period-radius relation
$R_2 \cong 7.9 \times 10^9 P_{\rm hr}~ {\rm cm}$
for the donor star.
Using these relations, we can get an appropriate estimation of the mass and radius of the secondary,
as $M_2 \simeq 0.36/0.33^{+0.07}_{-0.06}/0.43 M_{\odot}$ and $R_2 \simeq 0.41/0.41^{+0.12}_{-0.11}/0.45 R_{\odot}$,
corresponding to an M2-M3 dwarf.
The H and He emission features in the optical spectra \citep{Israel2016}
further leads to the conclusion that CXOGSG J2155 is an MCV.

Recently, \citet{Knigge2011} reconstructed the complete evolutionary path of CVs and presented
a full range of binary and donor properties.
With the assumption that the period 14101.10 s is due to the orbit,
we determine the binary properties from these revised (optimal) CV evolution models in \citet{Knigge2011}.
This leads to a binary system with the separation as 1.282 $R_{\odot}$,
containing a 0.313 $M_{\odot}$ (M3.5 type) secondary when assuming a 0.75 $M_{\odot}$ WD.
The physical properties for the secondary
are $R = 0.389 R_{\odot}$,
$T_{eff} = 3443$ K, log $g$ = 4.754 cm$^2$/s, and ${\dot{M}} = 10^{-8.872} M_{\odot}$/yr (mass loss rate).
The spectral type of the secondary is in good agreement with the
conclusion from the optical spectrum (Section \ref{ospec.sec}).


\citet{Knigge2011} listed model optical magnitudes of the secondary,
including $BVRI$ magnitudes on Johnson-Cousins system
and $JHK_{S}$ on CIT system, which can be transformed back to 2MASS system
using the equations given by \citet{Carpenter2001}.
Based on these absolute $JHK_{S}$ magnitudes,
which is only from the secondary and not affected by the primary,
the photometric distance is calculated as $64.2_{-2.27}^{+2.44}$ pc,
and the extinction is calculated as $E(B-V) = 0.42_{-0.117}^{+0.118}$.
This extinction is higher than the absorption by Galactic HI column density \citep{Kalberla2005} in this direction
($N_H = 1.65 \times 10^{21}$ cm$^{-2}$; $E(B-V) = 0.24$),
indicating an amount of internal absorption in the binary system.
The distance can be applied to obtain the X-ray luminosity as $L_X = 2\pi~f_{CEMEKL}D^{2} \simeq 1.5 \times 10^{28}$ erg s$^{-1}$.
Assuming this X-ray luminosity is entirely due to accretion, the accretion rate can be estimated from
$L_X = \frac {GM_{\rm WD} \dot{m}} {R_{\rm WD}}$.
Adopting the model parameters for the WD
$M_{\rm WD}$ = 0.75 $M_{\odot}$ and $R_{\rm WD}$ = 7.842 $\times$ 10$^8$ cm \citep{Knigge2011},
we obtain an accretion rate of $1.9 \times 10^{-15}$ $M_{\odot}$/yr.
The low X-ray luminosity and accretion rate may suggest CXOGSG J2155 as a low-accretion-rate polar
\citep[LARP;][]{Schwope2002a,Ramsay2004},
such as DP Leo \citep[$L_X \simeq 2.5 \times 10^{29}$ erg s$^{-1}$;][]{Schwope2002b},
WX LMi \citep[$\dot{m} \sim 1.5 \times 10^{-13}$ $M_{\odot}$/yr;][]{vogel2007},
EF Eri \citep[$L_X \simeq 2 \times 10^{29}$ erg s$^{-1}$;][]{Schwope2007},
and HU Aqu \citep[$L_X \simeq 4.7 \times 10^{28}$ erg s$^{-1}$;][]{Schwarz2009},
which is likely the progenitor of one polar \citep{Schmidt2005, Webbink2005,Kafka2010}.

It seems that there is a significant discrepancy between the mass-loss rate of the secondary
and the mass-accretion rate of the primary.
However, the mass-loss rate estimated by \citet{Knigge2011} is relevant to long timescales,
and it may be not applicable to some types of variations occurred on short timescales,
such as the low states of strongly magnetic polars with timescales of weeks to years
\citep[e.g.,][]{Hessman2000, Kafka2005, Knigge2011}.
On the other hand, for an LARP, the secondary is considered under-filling its Roche lobe
and the primary is accreting from the stellar wind of the secondary \citep{Schwope2002a, Vogel2011}.
That explains why LARPs are in a permanent low state and implies
that the wind mass-loss rate of the secondary might actually be of the same order of the mass-accretion rate.


3. ${Whether\ the\ radio\ counterpart\ is\ true?}$

Surveys of non-magnetic CVs \citep{Cordova1983, Fuerst1986} suggest that they are not radio emitters except in an outburst.
However, a number of MCVs have been confirmed as significant radio emitters, such as
AM Her \citep{Chanmugam1983}, V834 Cen \citep{Wright1988}, AE Aqr \citep{Bastian1988},
DQ Her \citep{Pavelin1994}, and AR UMa \citep{Mason2007}.
The flux densities of a few MCVs (within 200 pc) exceed 1 mJy, and flare events are present in several MCVs with the flux densities up to 10--30 mJy
\citep[see, e.g., a summary in][]{Mason2004}. Therefore, the radio intensity of CXOGSG J2155 is normal among MCVs.
The significance for CXOGSG J2155 is above 15$\sigma$ and 8$\sigma$ for the 4.91 GHz and 8.66 GHz observations.
However, the VLA images seriously suffer from the interference stripes, and some bright spots could be mistaken for true sources
due to a bad uv coverage and low sensitivity of the VLA data.
In addition, we notice that CXOGSG J2155 is less than two arcmin away from the radio galaxy 3C 438,
which is embedded in a massive galaxy cluster 1E 0657-56 \citep{Markevitch2002, Kraft2007}.
Therefore, the radio emission may be also due to background galaxies.
The radio counterpart, if real, suggests that the two observations (4.91 GHz and 8.66 GHz) correlate with radio flares.
More radio observations are needed to identify the radio counterpart and confirm the
properties of the CV binary.


\begin{acknowledgements}

This work has made use of data obtained from the Chandra Data Archive,
and software provided by the Chandra X-ray Center (CXC) in the application packages CIAO.
This research has made use of the SIMBAD database and the VizieR catalogue access tool,
operated at CDS, Strasbourg, France.
This research made use of Astropy,
a community-developed core Python package for Astronomy (Astropy Collaboration, 2013).
The authors acknowledge support from the National Science Foundation of China under grants
NSFC-11273028 and NSFC-11333004, and support from the National Astronomical Observatories,
Chinese Academy of Sciences under the Young Researcher Grant.

\end{acknowledgements}


\end{document}